# Magnetism of PrFeAsO parent compound for iron-based superconductors: Mössbauer spectroscopy study

K. Komędera[1], A. Pierzga[1], A. Błachowski[1], K. Ruebenbauer[1*], A. Budziak[2], S. Katrych[3], and J. Karpinski[3]

[1]Mössbauer Spectroscopy Division, Institute of Physics, Pedagogical University
*PL-30-084 Kraków, ul. Podchorążych 2, Poland*

[2]Institute of Nuclear Physics, Polish Academy of Science
*PL-31-342 Kraków, ul. Radzikowskiego 152, Poland*

[3]Institute of Condensed Matter Physics, EPFL
*CH-1015 Lausanne, Switzerland*

[*]Corresponding author: sfrueben@cyf-kr.edu.pl

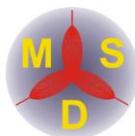

## Abstract

Mössbauer spectroscopy measurements were performed for the temperature range between 4.2 K and 300 K in a transmission geometry applying 14.41-keV resonant line in $^{57}$Fe for PrFeAsO the latter being a parent compound of the iron-based superconductors belonging to the '1111' family. It was found that an itinerant 3d magnetic order develops at about 165 K and it is accompanied by an orthorhombic distortion of the chemical unit cell. A complete longitudinal 3d incommensurate spin density wave (SDW) order develops at about 140 K. Transferred hyperfine magnetic field generated by the praseodymium magnetic order on iron nuclei is seen at 12.8 K and below, i.e., below magnetic order of praseodymium magnetic moments. It is oriented perpendicular to the field of SDW on iron nuclei. The shape of SDW is almost rectangular at low temperatures and it transforms into roughly triangular form around "nematic" transition at about 140 K. Praseodymium magnetic order leads to the substantial enhancement of SDW due to the large orbital contribution to the magnetic moment of praseodymium. A transferred field indicates presence of strong magnetic susceptibility anisotropy in the [b-c] plane while following rotation of praseodymium magnetic moments in this plane with lowering temperature. It was found that "nematic" phase region is a region of incoherent spin density wavelets typical for a critical region.

## 1. Introduction

PrFeAsO is one of the parent compounds for various iron-based superconductors belonging to the major '1111' family with the highest bulk transition temperature to the superconducting state [1-7]. Superconductors could be obtained from the parent compound by either electronic, hole or isovalent replacement of any element by another suitable element. Similar effects could be achieved applying hydrostatic pressure. High superconducting transition temperatures are assured by the presence of strongly internally coupled corrugated "two-dimensional" iron-pnictogen layers separated by large distances one from another thanks to the complex and rather thick rare earth – oxygen layers. Fe-As layers are stacked in the normal order without inversion [8-10]. Compounds containing praseodymium as rare earth are particularly interesting due to the large localized magnetic moment of praseodymium with significant orbital contribution. In fact, replacement of e. g. lanthanum by the rare earth with the large magnetic moment having significant orbital contribution like Pr or Nd leads to the increase of the superconducting transition temperature [3, 8]. The parent compounds of the iron-based superconductors behave like metals. Parents of the major '1111' family crystallize in the tetragonal structure with *P4/nmm* symmetry [11]. They undergo orthorhombic distortion upon lowering temperature followed by the antiferromagnetic order development of the 3d itinerant type. Namely, a longitudinal spin density wave (SDW) develops along one of the main axes within former tetragonal plane [12-14]. The SDW has complex shape evolving with temperature and it is incommensurate with the corresponding crystal lattice periodicity. A development of the itinerant magnetism and orthorhombic distortion are suppressed within superconductor, i.e. one observes 3d diamagnetic behavior. On the other hand, localized magnetic moments of the rare earth order magnetically at low temperatures even within superconducting phase [15-18]. This order has antiferromagnetic symmetry. One can obtain overdoped regime for majority of iron-based superconductors with neither superconductivity nor 3d magnetic moments, albeit still exhibiting metallic behavior. For PrFeAsO parent compound one observes orthorhombic distortion at 165 K accompanied with some incoherent itinerant 3d magnetism. This so-called "nematic" phase survives till about 140 K (see, Ref. [9]), and at this temperature a development of the coherent SDW begins. Praseodymium orders antiferromagnetically within practically saturated SDW phase at about 12 K. Localized magnetic moments of praseodymium are oriented perpendicular to the Fe-As planes. A partial rotation of the praseodymium moments on the former tetragonal plane occurs at lower temperatures and it seems to be completed above 4.2 K. Antiferromagnetic order of praseodymium moments is preserved during this rotation [19-22]. One can expect small almost axially symmetric electric field gradient tensor on iron nuclei with the principal component being perpendicular to the Fe-As plane. Some early $^{57}$Fe Mössbauer data were obtained for PrFeAsO compound in rough agreement with statements made above [13]. However, it seems interesting to undertake some more detailed studies of this complex system by means of the $^{57}$Fe Mössbauer spectroscopy applying 14.41-keV resonant transition. It is well known that Mössbauer spectroscopy is very useful in investigation of complex magnetic structures [23-25] and interplay between magnetism and superconductivity [26, 27].

## 2. Experimental

The sample was synthesized from powders of PrAs and FeO (molar ratio 1/1). Both components were well mixed and pressed into pellets. The sample was heated in evacuated quartz tube during 17 h up to the 1050 °C and kept at this temperature for one week and finally quenched in cold water.

Mössbauer spectra have been collected in standard transmission geometry for 14.41-keV transition in $^{57}$Fe by using commercial $^{57}$Co(Rh) source kept under ambient pressure and at room temperature. Absorber was made in the powder form mixing 44 mg of PrFeAsO with the B$_4$C carrier. Absorber thickness amounted to 21.9 mg/cm$^2$ of PrFeAsO, the latter having natural isotopic composition. A Janis Research Co. SVT-400 cryostat was used to maintain the absorber temperature, with the long time accuracy better than 0.01 K (except for 4.2 K, where the accuracy was better than 0.1 K). A RENON MsAa-3 Mössbauer spectrometer equipped with a Kr-filled proportional counter was used to collect spectra in the photo-peak window. Velocity scale of the Mössbauer spectrometer was calibrated applying Michelson-Morley interferometer equipped with the He-Ne laser. Spectral shifts are reported versus ambient pressure and room temperature natural α-Fe. Spectra were fitted within transmission integral approximation by means of the Mosgraf-2009 applications [29].

## 3. Results and discussion

Powder X-ray diffraction pattern of the PrFeAsO sample used in the further research is shown in Figure 1. The pattern was obtained by using Cu-Kα radiation and the structure of the main phase was refined within *P4/nmm* (*Z*=2) space group obtaining the following lattice constants $a = 0.39836(1)$ nm and $c = 0.86103(3)$ nm at room temperature. Note presence of some contamination by praseodymium oxide being most likely Pr$_6$O$_{11}$ – fortunately free of iron [28]. Unfortunately, crystallographic information, which could help quantitatively estimate the phase ratio in the sample, is unknown [28].

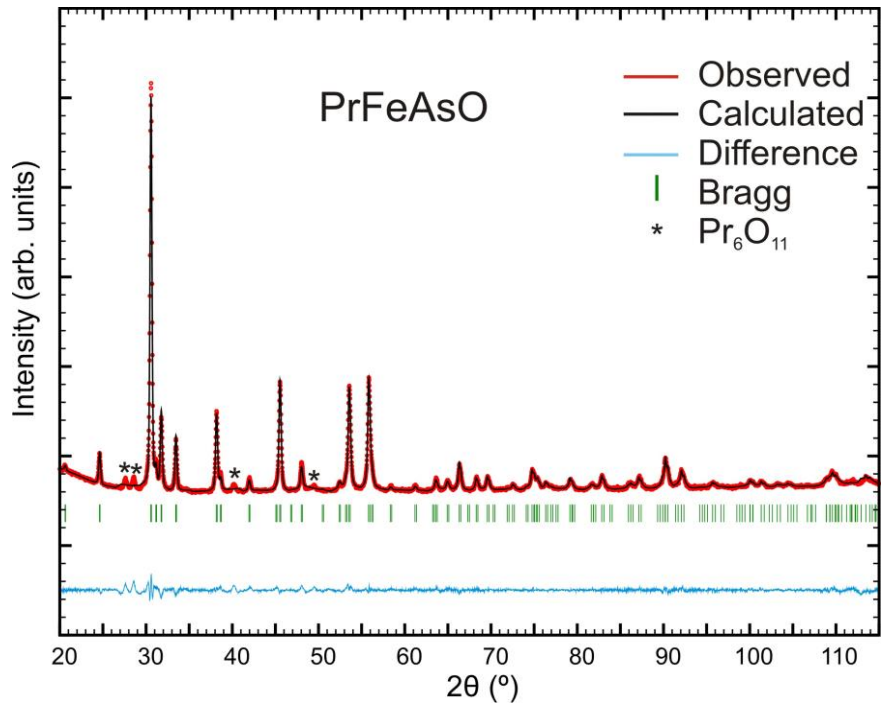

**Figure 1** Powder X-ray diffraction pattern plotted versus scattering angle 2θ obtained at room temperature for PrFeAsO sample – see text for more details.

Mössbauer spectra obtained above magnetic 3d ordering and within "nematic" phase are shown in Figure 2.

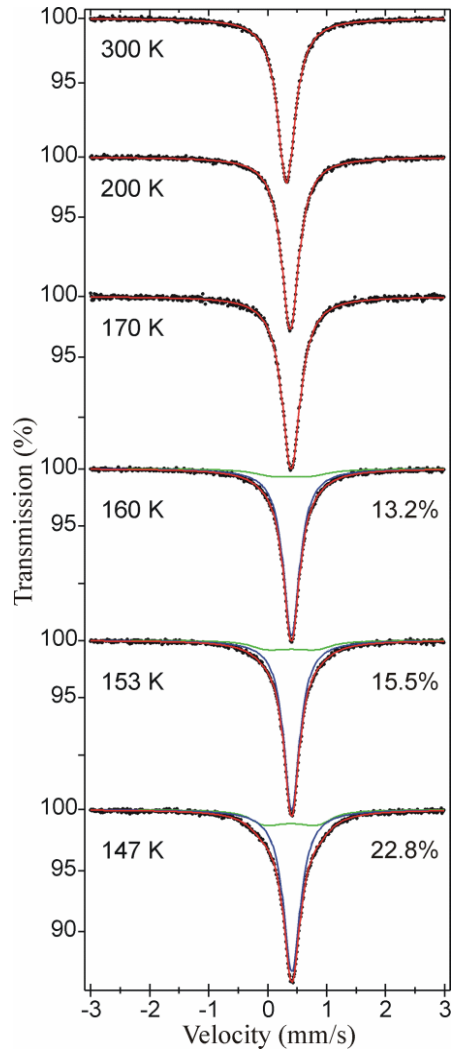

**Figure 2** $^{57}$Fe Mössbauer spectra of PrFeAsO versus temperature within non-magnetic and "nematic" regions. For "nematic" region relative contribution of the magnetically ordered phase is shown together with two "sub-spectra" resulting from the non-magnetic and magnetic components, respectively.

One can see that spectra are characterized by quite well defined singlet above magnetic ordering temperature. Upon development of the "nematic" phase the majority of spectrum remains as unsplit singlet, albeit some component with the magnetic dipole interaction and electric quadrupole interaction occurs. A contribution of this component increases gradually with lowering of the temperature. These results are consistent with the previous findings that magnetic order starts at about 165 K, and that "nematic" phase survives till about 140 K down the temperature scale [9, 19-21]. It is interesting to note, that the electric field gradient (EFG) tensor on iron nuclei is below detection threshold for this tetragonal phase.

Figure 3 shows Mössbauer spectra obtained within SDW region, albeit above magnetic ordering of praseodymium. Corresponding shapes of SDW are shown together with respective normalized hyperfine magnetic fields distributions. These spectra are obtained below crystallographic transition to the orthorhombic phase.

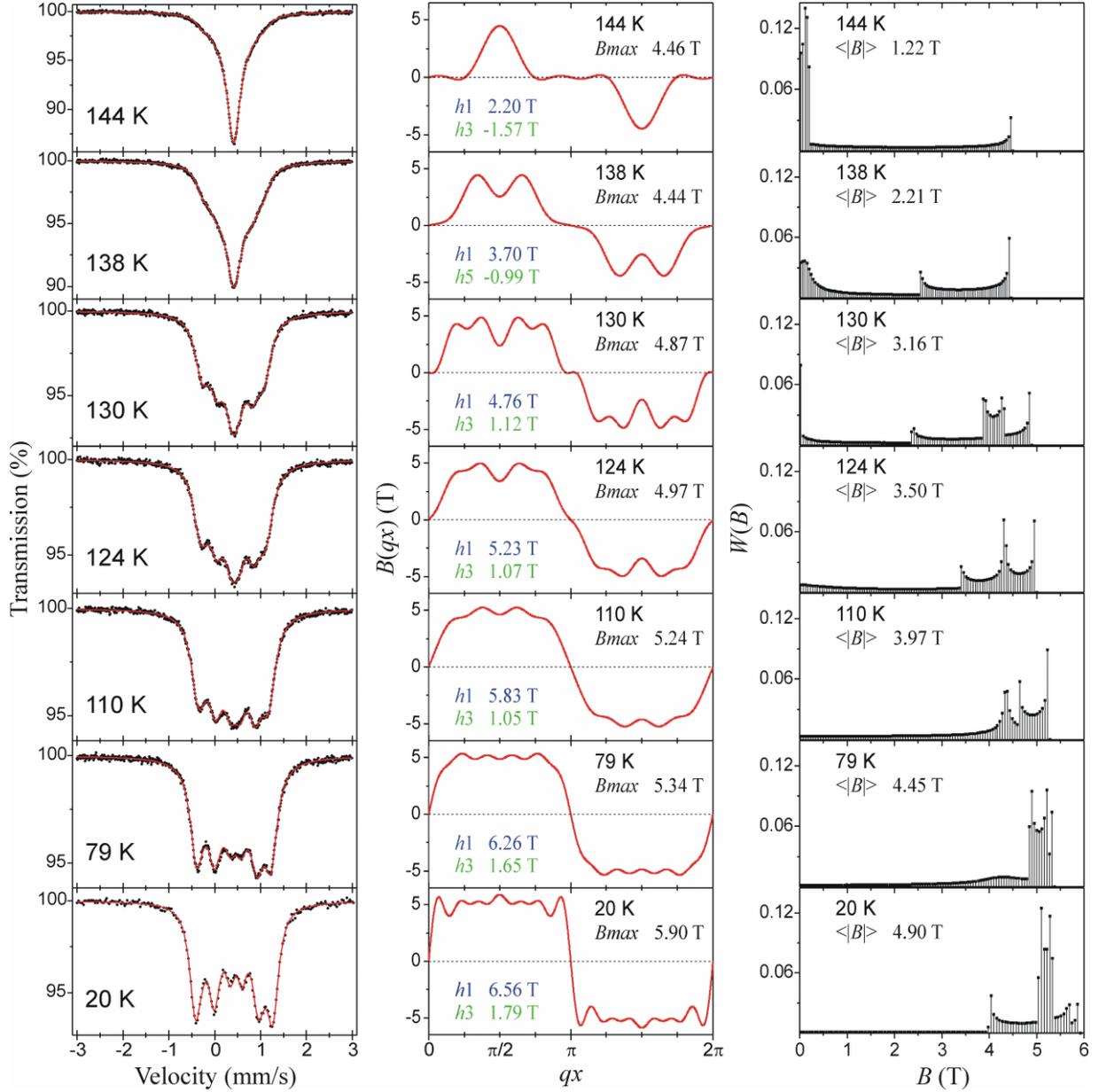

**Figure 3** $^{57}$Fe Mössbauer spectra of PrFeAsO versus temperature within SDW region, albeit above magnetic ordering of praseodymium (left column). Central column shows corresponding shapes of SDW $B(qx)$ versus phase angle $qx$. The symbol $B_{max}$ stands for the maximum absolute amplitude of SDW. Amplitudes of the first two dominant harmonics are show as well. Resulting distributions $W(B)$ of the magnetic hyperfine field $B$ are shown in the right column. The symbol $\langle |B| \rangle$ denotes average field.

The shape of SDW evolves in a typical way for parents of the iron-based superconductors during lowering of the temperature. It has nearly triangular shape with large regions devoid of the field at high temperature, and eventually evolves to the nearly rectangular shape close to saturation [30]. Some small EFG tensor appears on iron nuclei within this region. A quadrupole interaction was treated in the first order approximation and under assumption that it has axial symmetry (due to its smallness). Essential parameters of SDW harmonics are gathered in Table 1. Details concerned with the SDW processing could be found as section 3 of Ref. [30]. Evaluation of spectra exhibiting SDW and/or transferred field due to praseodymium magnetic order was performed by using *GmfpHARM* application of the Mosgraf-2009 Mössbauer data processing suite [29]. Transferred field could be fitted as two components one parallel to the SDW field and another one as perpendicular. Hence, one obtains the following total hyperfine magnetic field $B_T = \sqrt{B_{perp}^2 + [B(qx) + B_{par}]^2}$ with the symbol $B_{perp}$ denoting component of the transferred field perpendicular to the SDW field $B(qx)$ and the symbol $B_{par}$ denoting component of the transferred field parallel to the SDW field. A field distribution is created solely by the SDW field, of course.

**Table 1**

Amplitudes of subsequent harmonics $h_n$ ($n = 1, 3, ...$) used to evaluate SDW shape for several spectra obtained at various temperatures $T$. The first harmonic $h_1$ has positive amplitude by definition [30].

| $T$ (K) | $h_1$ (T) | $h_3$ (T) | $h_5$ (T) | $h_7$ (T) | $h_9$ (T) | $h_{11}$ (T) | $h_{13}$ (T) |
|---|---|---|---|---|---|---|---|
| 144 | 2.23(3) | -1.58(5) | 0.65(3) | - | - | - | - |
| 138 | 3.58(7) | -0.42(9) | -1.1(2) | 0.3(1) | - | - | - |
| 130 | 4.75(4) | 1.12(8) | -0.48(6) | -0.04(5) | -0.7(1) | - | - |
| 124 | 5.23(2) | 1.08(5) | -0.28(2) | 0.09(2) | -0.38(5) | - | - |
| 110 | 5.82(1) | 1.04(8) | 0.2(1) | 0.2(1) | -0.15(2) | - | - |
| 79 | 6.25(1) | 1.63(6) | 0.57(1) | 0.17(6) | 0.1(1) | 0.26(4) | - |
| 20 | 6.56(3) | 1.79(6) | 1.14(6) | 0.78(2) | 0.1(1) | 0.59(4) | 0.44(5) |
| 12.8 | 6.87(2) | 1.63(7) | 1.52(6) | 0.60(2) | 1.17(6) | 0.51(5) | 0.72(4) |
| 10.6 | 8.00(5) | 2.5(1) | 1.90(4) | 0.63(2) | 0.23(5) | 0.60(6) | 0.39(5) |
| 8.4 | 8.29(7) | 2.70(5) | 1.80(4) | 0.63(2) | 0.23(5) | 0.60(6) | 0.39(5) |
| 6.2 | 8.29(4) | 2.69(7) | 1.76(4) | 0.63(2) | 0.23(5) | 0.60(6) | 0.39(5) |
| 4.2 | 8.41(4) | 2.74(5) | 1.68(3) | 0.63(2) | 0.23(5) | 0.60(6) | 0.39(5) |

Figure 4 shows Mössbauer spectra across magnetic ordering of the praseodymium localized magnetic moments.

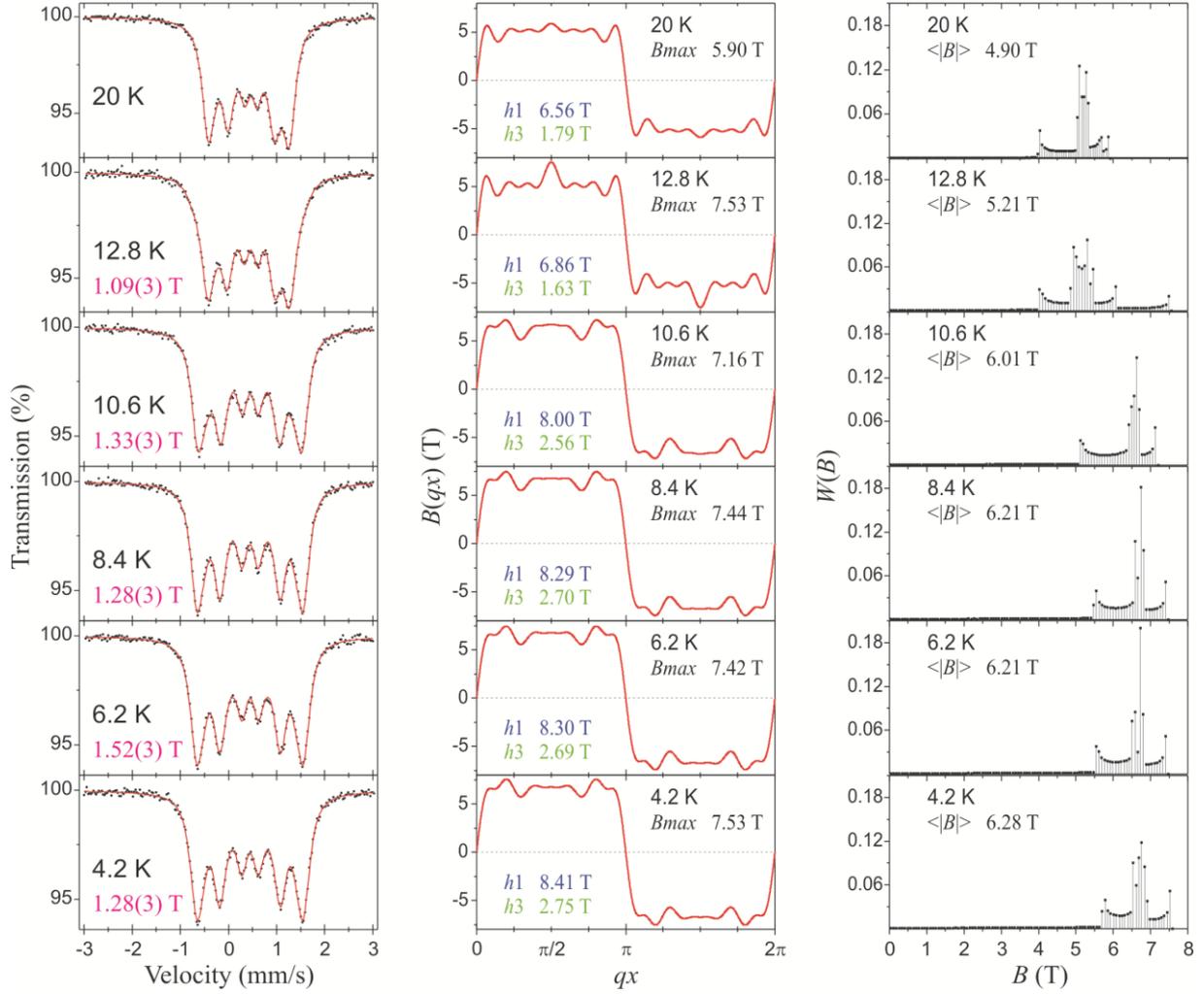

**Figure 4** $^{57}$Fe Mössbauer spectra of PrFeAsO across magnetic ordering of localized praseodymium magnetic moments (left column). Transferred hyperfine magnetic fields on iron nuclei due to the praseodymium order are shown as well. Central column shows resulting shapes of SDW, while the right column corresponding field distributions due to SDW. All symbols of the central and right column have the same meaning as for Figure 3. Plots for 20 K are taken from Figure 3.

The best fits to the data are obtained with a transferred hyperfine magnetic field on iron nuclei being perpendicular to the SDW field, the latter being oriented along one of the previous tetragonal axes. There is virtually no "longitudinal" transferred field, but the amplitude of SDW is significantly enhanced due to the praseodymium magnetic order. One has to note that SDW at 20 K is practically saturated. A transferred field appears at 12.8 K in fair agreement with the previously found ordering temperature of praseodymium (12 K). A transferred field seems to have maximum at about 6.2 K. One can conclude that the overall magnetic system is already saturated at 4.2 K. There is no other effect of the praseodymium magnetic moments rotation on the Mössbauer spectra except above apparent maximum of the transferred field. Hence, one can conclude that rotation occurs in the plane perpendicular to the propagation direction of the longitudinal SDW.

It is interesting to note that apparent electric quadrupole interaction on iron nuclei gradually vanishes with praseodymium ordering despite lack of any crystallographic transition within this temperature region. The EFG tensor is likely to be axially symmetric with the principal axis oriented perpendicular to the Fe-As plane. Hence, this axis is perpendicular to the SDW field and SDW propagation direction, as SDW is longitudinal and propagates along one of the former tetragonal main axis aligned with the Fe-As plane. One can take into account effect of the transferred field leading to the so-called "magic angle" conditions. However, it is definitely insufficient to explain observed effect, particularly taking into consideration rotation of the praseodymium moments in contrast to the claims of Ref. [13]. It is rather likely that higher charge-symmetry is restored by the magneto-elastic effects. These effects could be caused by significant orbital contribution to the localized magnetic moment of praseodymium.

Figure 5 shows essential Mössbauer parameters versus temperature $T$, i.e., a total central shift (CS) denoted as $S$ versus room temperature α-Fe, effective quadrupole coupling constant $A_Q = \frac{1}{24} eQ_e V_{zz} (c/E_0)(3\cos^2\beta - 1)$, and the average magnetic field of SDW $\langle |B| \rangle$. It is assumed that the EFG tensor is axially symmetric with the principal axis making angle $\beta$ with the hyperfine magnetic field. The symbol $e$ stands for the positive elementary charge. The symbol $Q_e$ denotes spectroscopic electric quadrupole nuclear moment in the first excited state of $^{57}$Fe, the symbol $V_{zz}$ stands for the principal component of the EFG, the symbol $c$ denotes speed of light in vacuum, and the symbol $E_0$ stands for the resonant transition energy. In the absence of the hyperfine field one assumes that the following condition is satisfied $\beta = 0$.

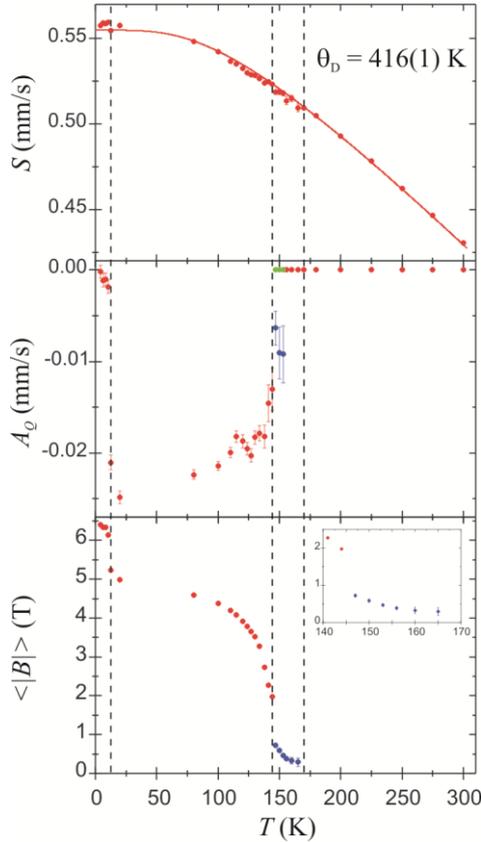

**Figure 5** Essential Mössbauer parameters for PrFeAsO plotted versus temperature $T$. Vertical dashed lines are borders of different regions. The rightmost region is a region without magnetic order. The next region is a region of "nematic" phase followed by the region of SDW order without praseodymium order. The leftmost narrow region is a region of SDW and praseodymium order. The top part shows the total central shift (CS) denoted as $S$ relative to the total shift in room temperature α-Fe. The variation versus temperature is entirely due to the second-order Doppler shift (SOD) and fit (solid line) yields Debye temperature $\theta_D = 416(1)$ K. The central part shows effective quadrupole coupling constant $A_Q$. Green points correpond to the non-magnetic component of the "nematic" phase, while the navy-blue to the magnetic component of this phase. The lowest part shows average SDW field $\langle |B| \rangle$. Navy-blue points show average field on iron nuclei within "nematic" phase. Inset shows fields on expanded temperature scale around "nematic" transition.

It is obvious that lattice dynamics is insensitive to all transitions involved, as the second order Doppler shift (SOD) behaves regularly all the way from the room temperature till the ground state of the system. Hence, the magneto-elastic forces are weak as expected for highly itinerant magnetism. It is assumed that the isomer shift remains constant within this temperature region, as this region is described by the harmonic atomic motions, and hence it exhibits negligible thermal expansion. A Debye temperature is typical for the strongly bound metal-covalent system. An orthorhombic distortion does not change electron density on the iron nuclei for this '1111' system with widely spaced Fe-As layers. A negative effective quadrupole constant is an indication that in the orthorhombic phase with 3d itinerant magnetic order, albeit with disordered magnetically praseodymium one has small (axially symmetric) EFG with the principal component being positive and oriented perpendicular to the Fe-As plane. This component gradually increases with lowering of the temperature. Praseodymium magnetic order restores higher charge symmetry on iron nuclei and the quadrupole interaction vanishes.

Figure 6 shows in more detail the average SDW field $\langle|B|\rangle$ versus temperature $T$ including average field on iron nuclei in the "nematic" phase.

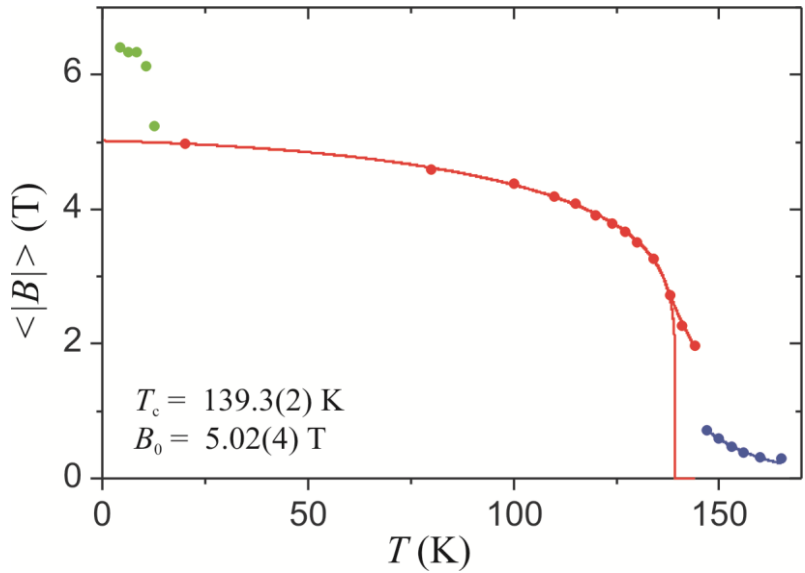

**Figure 6** Average field due to SDW on iron nuclei $\langle|B|\rangle$ plotted versus temperature $T$. Green points show enhancement due to the magnetic order of praseodymium. Blue points show average field in the "nematic" phase (solid blue line is a fit to the power law). Red points represent coherent SDW region except short partly incoherent tail with inverted curvature. Solid red lines represent fit yielding transition temperature $T_c = 139.3(2)\,\mathrm{K}$ and saturation field (excluding enhancement) $B_0 = 5.02(4)\,\mathrm{T}$.

One obtains typical saturation field of SDW excluding enhancement due to the praseodymium magnetic order as far as parents of the iron-based superconductors are considered. A transition temperature to the coherent SDW region is in perfect agreement with previous reports [19-21]. The so-called "nematic" phase seems to be a region of incoherent SDW with the large part of the sample being already in the non-magnetic state. Some islands of the magnetically ordered phase survive till quite high temperatures and they seem to remain in the orthorhombic phase, while non-magnetic parts are already in a tetragonal phase. Such behavior is peculiar for the compound with praseodymium as the rare earth.

## 4. Conclusions

There are four distinct temperature regions for PrFeAsO. Namely, a complete magnetic disorder and tetragonal phase above 165 K. A partial 3d itinerant magnetic order in the form of incoherent SDW below 165 K and above about 140 K. For higher temperatures within this region one can observe some "magnetic islands" already in the orthorhombic phase. However, it is hard to get electric quadrupole interaction parameters close to the completely non-magnetic region, as the amount of magnetic phase is very low in this region. Upon lowering temperature one obtains almost pure orthorhombic phase with incoherent longitudinal SDW. The latter transition is quite sharp (probably of the first order) leading to a jump of the average magnetic field on the iron nuclei. This uppermost region with magnetic order traces is called "nematic" phase, as the magnetic 3d itinerant order seems to be coupled to the orthorhombic distortion in this material. The long range coherent SDW occurs below $T_c = 139.3(2)$ K in the orthorhombic phase. SDW appears in the triangular form with large proportions of the almost magnetism-free regions and evolve to the almost rectangular shape at low temperature. The saturation occurs well above praseodymium magnetic order. Above behavior shows close similarity to the behavior of the '122' compound $BaFe_{2-x}Ru_xAs_2$ with the iron content high enough to have 3d magnetic order [31].

One observes small axially symmetric EFG tensor on iron nuclei in the orthorhombic phase with the principal axis being perpendicular to the Fe-As plane and with the positive principal component. This component increases with lowering of the temperature. On the other hand, the EFG tensor on iron nuclei in the tetragonal phase is below detectability limit.

Praseodymium orders antiferromagnetically at about 12.8 K with the localized magnetic moments being perpendicular to the Fe-As planes. This is the fourth and lowest temperature region. A transferred field is observed on iron nuclei increasing till saturation of the praseodymium magnetism. This field is perpendicular to the SDW field. On the other hand, praseodymium magnetic order enhances amplitude of SDW. Hence, it has some effect on the spin polarization of the itinerant electrons involved in 3d magnetism. A rotation of the praseodymium ordered magnetic moments was reported above liquid helium temperature [4, 19, 21]. It has no influence on the iron Mössbauer spectra except apparent maximum of the hyperfine transferred field at about 6.2 K. Hence, a rotation occurs in the plane perpendicular to the propagation direction of longitudinal SDW. It seems that transferred field saturates quite rapidly with lowering of the temperature. An apparent maximum of this field might be caused by high anisotropy of the magnetic susceptibility in the planes perpendicular to the Fe-As plane, as the system is highly planar. Such finding suggests that it is much easier to polarize magnetically itinerant electrons by the magnetic field perpendicular to the Fe-As plane than by the in-plane field. It seems that transferred field is aligned with the praseodymium magnetic moment exhibiting significant orbital contribution.

The electric quadrupole interaction on iron nuclei vanishes with the magnetic order of praseodymium without change in the crystal symmetry. It has to be some magneto-elastic effect due to the large orbital contribution to the praseodymium magnetic moment. One cannot explain this phenomenon by the "magic angle" effect as stated in Ref. [13], as the angle between principal axis of the EFG and magnetic hyperfine field diminishes from the right angle to about ~76$^o$ at most. This means that the effective quadrupole coupling constant is reduced in absolute terms by about 18 % only under such circumstances.